\documentclass[conference]{IEEEtran}
\usepackage{amsfonts}
\usepackage{mathbbol}
\usepackage{amssymb}

\ifCLASSINFOpdf
\else
\fi
\usepackage{indentfirst}
\usepackage[T1]{fontenc}

\usepackage{mathrsfs}

\usepackage{amsmath}

\usepackage{graphicx}
\usepackage{epstopdf}

\usepackage{booktabs}

\usepackage{color}

\usepackage{algorithm}
\usepackage{algorithmic}

\usepackage{cite}

\begin{document}
%
\title{ {\color{black}Joint Domain Based} Massive Access for Small Packets Traffic of Uplink Wireless Channel}
\author{\IEEEauthorblockN{Qiang Song, Ronggui Xie, Huarui Yin and Guo  Wei}
%
\IEEEauthorblockA{School of information Science and Technology\\
University of Science and Technology of China, Hefei, Anhui 230027\\
Email: \{icesong, xrghit03\}@mail.ustc.edu.cn, \{yhr,wei\}@ustc.edu.cn}
}


\maketitle

\begin{abstract}
The fifth generation (5G) communication scenarios such as the cellular network and the emerging machine type communications will produce massive small packets.
To support massive connectivity and avoid signaling overhead caused by the transmission of those small packets,
this paper proposes a novel method to improve the transmission efficiency for massive connections of wireless uplink channel.
The proposed method combines {\color{black}compressive sensing (CS)} with power domain NOMA jointly,
especially neither the scheduling nor the centralized power allocation is necessary in the method.
Both the analysis and simulation show that the method can support up to two or three times overloading.

\end{abstract}

%
\IEEEpeerreviewmaketitle

\section{Introduction}
The fifth generation (5G) system is supposed to provide high quality communication service for a series of scenarios, for example, enhanced mobile broadband and massive machine type communications\cite{3gpp2017study}. Massive terminals in these scenarios will produce huge data traffic. Besides, the data packets to {be delivered} are usually very short, i.e., small packets\cite{grinter2002instant,xiao2007understanding}, especially in the uplink.
%
In order to offer higher spectrum efficiency and support massive connectivity, 
non-orthogonal multiple access (NOMA) schemes have attracted great interest, both from the academia and industry\cite{Benjebbour2013Concept, Sun2015On}.

Typical NOMA schemes include sparse code multiple access (SCMA)\cite{nikopour2013sparse}, multi-user shared access (MUSA)\cite{yuan2016multi} and power domain NOMA\cite{islam2016power}, etc.
Both SCMA and MUSA achieve multiplexing in code domain, while power domain NOMA attains multiplexing in power domain. All of them can realize multi-user detection (MUD) and overloading.
 However, the SCMA  faces the challenges of high-dimensional code design and exponential growth of recovery complexity, and the MUSA faces the challenge of improving the recovery accuracy.
Compared with code domain NOMA, power domain NOMA usually has a lower complexity as the number of users increases\cite{dai2015non}.
There are many works for downlink power domain NOMA\cite{saito2013non, ding2014performance, yu2004iterative}, etc.
While in this paper we pay more attention to uplink power domain NOMA.
Ding et.al proposed a general NOMA framework for both downlink and uplink transmission by applying the concept of signal alignment in \cite{ding2016general},
but it focused on power allocation for the two-users pair, and it is not applicable for massive connectivity.
Zhang et al. proposed  an uplink power control scheme in \cite{zhang2016uplink} with center controlled and back-off scheme, thus signaling can not be avoidable for scheduling, which introduces the large latency at the same time.

In the services of small packet, especially for grant-free uplink transmission, there is an issue that the users' activity or instantaneous system loading is not readily known to the receiver.
Compressive sensing (CS)\cite{donoho2006compressed} is a promising technique to estimation parse signals and support MUD
\cite{bockelmann2013compressive}. 
\cite{xie2016many} proposed a compressive sensing based (CS-based) massive access
{\color{black}scheme} for small packets for uplink wireless channel, which is simply referred to herein as SPMA. The system is modeled as a block-sparsity pattern with unknown sparsity level. Block precoding is employed at each single-antenna user equipment (UE), and interference cancellation based block orthogonal matching pursuit (ICBOMP) algorithm is developed at the base station (BS).
Simulation results demonstrated the effectiveness of the CS-based scheme in small packet services.
{\color{black}We should mention that this CS-based method can also be regarded as a special code-based access scheme, which spreads each symbol to whole time-frequency resource with dedicated precoding vector.}
However, the overloading is not considered in SPMA.

In this paper, combining CS with power domain NOMA, we propose a joint domain based massive access for small packet traffic of uplink wireless channel, named as JSPMA later. In the proposed algorithm, for users, the messages are precoded by different precoding matrices. Besides, users gather into several groups based on their own channel amplitudes, users in the same group transmit messages with the same power, while the transmission power of adjacent groups are gradually degraded with a certain step. For BS, it detects and distinguishes users based on ICBOMP algorithm and then carries out successive interference cancellation (SIC) for next detection. The proposed method can greatly enlarge overloading factor and reduce the signaling overhead for uplink many-access system.


\textbf{Notation}: Vectors and matrices are denoted by boldface lowercase and uppercase letters, respectively. The 2-norm of a column vector $\textbf{v}$ is denoted $\|\textbf{v}\|_2$, which equals to $\sqrt{\textbf{v}^H\textbf{v}}$, and $\textbf{v}^H$ denotes its conjugate transpose. For a subset $\emph{I} \subset [N] := {1,2,\cdots,N}$ and matrix $\textbf{A} := [\textbf{A}_1, \textbf{A}_2, \cdots, \textbf{A}_N] $ consisting of $N$ sub-matrices with equal size, where $:=$ means definition. $\textbf{A}_\emph{I}$ stands for sub-matrices of $\textbf{A}$ whose block indices are in $\emph{I}$. Given two sets $\emph{I}_1$ and $\emph{I}_2$, $\emph{I}_1 \setminus \emph{I}_2 := \emph{I}_1 \cap \emph{I}_2^c$. The Kronecker product $\otimes$ is required with the use of the column stacking operator $\text{vec}(\cdot)$, and $\tbinom{m}{k}$ is the binomial coefficient for $m \ge k$.

\section{System Model}
We consider an uplink system with $N$ mobile users, each with a single antenna, and a BS with \(M\) antennas. When a terminal is admitted to the network, it becomes an \emph{online} user. We assume that there are $N_a$ \emph{active} users, out of the total $N$ online users, transmitting their data to the BS simultaneously. It is not required $N_a$ be known a priori or $N_a\textless M$. Actually in practical systems, $N_a$ is usually unknown and it is possible that $N_a \geq M$.


Since the lengths of small packets are usually shorter than $T$, the number of symbols of every frame, we extend their lengths by precoding to fully utilize the available resources. Let $ \textbf{s}_n \in \mathbb{C}^{d \times 1} $ denote the symbol vector to be transmitted by user $n$ ($U_n$), with $d \textless T$ ($n=1,2,\cdots,N$). User $n$  applies a precoding to $\textbf{s}_n$ to yield,
\begin{equation}
\textbf{c}_n=\textbf{P}_n\textbf{s}_n
\end{equation}
where $\bf{P}_n$ is a $ T \times d$ complex precoding matrix. {\color{black}We assume that $\bf{P}_n$ is normalized such that each column has 2-norm equal to 1}.

The $n$-th user will send data $\textbf{c}_n$ with transmission power ${\rho}_n $, so the data sent by the $n$-th user is expressed as
\begin{equation}
\textbf{x}_n=\sqrt{\rho_n}\textbf{P}_n\textbf{s}_n
\end{equation}

The observation at the BS is given by
\begin{equation}
\textbf{Y}=\sum_{n=1}^N\textbf{h}_n\textbf{x}_n^T+\textbf{Z}=\sum_{n=1}^N\textbf{h}_n\sqrt{\rho_n}\textbf{s}_n^T\textbf{P}_n^T+\textbf{Z}
\end{equation}
where $\textbf{Y}$ is the noisy measurement of size $M \times T$,
$\textbf{h}_n \in \mathbb{C}^{M \times 1}$ represents the complex channel response from the $n$-th user towards to the BS.  $\textbf{Z}\in\mathbb{C}^{M \times T}$ stands for the additive Gaussian noise matrix. Without loss of generality, each element of $\textbf{h}_n$, $\textbf{s}_n$ and $\textbf{Z}$ is assumed to be a zero-mean and unit-variance variable.

Applying the column stacking operator on both sides of above equation,
\begin{equation}
\text{vec}(\textbf{Y})=\sum_{n=1}^N(\textbf{P}_n \otimes \textbf{h}_n)\sqrt{\rho_n}\textbf{s}_n+ \text{vec}(\textbf{Z})
\label{Kronecker}
\end{equation}

Define $\textbf{y}:=\text{vec}(\textbf{Y})$, $\textbf{z}:=\text{vec}(\textbf{Z})$,  $\textbf{B}_n:=\textbf{P}_n \otimes \textbf{h}_n $, $\textbf{B}:=[\textbf{B}_1, \textbf{B}_2, \cdots , \textbf{B}_N]$ ,
$\textbf{s}:=[\textbf{s}_1^T,  \textbf{s}_2^T,  \cdots,  \textbf{s}_N^T]^T $ and
$\textbf{s}_{\rho}:=[\sqrt{\rho_1}\textbf{s}_1^T,  \sqrt{\rho_2}\textbf{s}_2^T,  \cdots,  \sqrt{\rho_N}\textbf{s}_N^T]^T$. Then (\ref{Kronecker}) can be rewrite as
\begin{equation}
\textbf{y}=\textbf{B}\textbf{s}_{\rho}+\textbf{z}
\label{receivedY}
\end{equation}

In our method, each UE determines its transmission power according to its own channel amplitude information, respectively.
The BS, as in \cite{xie2016many}, will recover the small packets by ICBOMP algorithm. The ICBOMP is developed from the well known sparse recovery algorithm BOMP. It takes use of the error correction and symbol detection of the communication packets to improve its recovery performance. Readers can find more details in \cite{xie2016many}.

$\textbf{Remark 1}$: when $\rho_1 = \rho_2 = \cdots = \rho_n = \rho_0$, $n \in [N_a]$, where $\rho_0$ is the mean of transmission power, the proposed model {\color{black}reduces to} the original SPMA.


\section{Power Allocation Policy}
\subsection{power domain NOMA in uplink}
Without loss of generality, we assume $\emph{I}_a = [N_a]$ and $\|\textbf{h}_1\|^2 \geq \|\textbf{h}_2\|^2 \geq \large{\cdots} \geq \|\textbf{h}_n\|^2 \geq \large{\cdots} \geq \|\textbf{h}_N\|^2$. Indices from $N_a+1$ to $\emph{N}$ are for the inactive users. For every user $n$, the \emph{unordered} random variable $\textbf{h}_n^H\textbf{h}_n$ follows a chi-squared distribute with $2M$ degrees of freedom. The probability density function (PDF) $f(x)$ and cumulative distribution function (CDF) $F(x)$ of unordered $\textbf{h}_n^H\textbf{h}_n$ are given as follows,
\begin{equation}
f(x)=\text{exp}(-x)\frac{x^{M-1}}{(M-1)!}
\label{chiPDF}
\end{equation}
\begin{equation}
F(x)=1- \text{exp}(-x)\sum_{k=0}^{M-1}\frac{x^{k}}{k!}
\end{equation}
both of which hold for $x \ge 0$, and are zero when $x<0$. By order statistics, the PDF of the \emph{ordered} $\textbf{h}_n^H\textbf{h}_n$ ($n=1,2,\cdots,N_a$) is given as
\begin{equation}
f_n(x)=n\tbinom{N_a}{n}F^{N_a-n}(x)[1-F(x)]^{n-1}f(x)
\label{order}
\end{equation}
the mean $\mathbb{E}\{  \textbf{h}_n^H\textbf{h}_n \}$ can be computed based on the above PDF.

During iteration, we can simply assume that the first $N_a$ iterators select all the $N_a$ active users and the selection order is based on the descending order of their $\textbf{h}_n^H\textbf{h}_n$. To keep order statistics, $\rho_n$ needs meet following condition generally (Proof is omitted here due to space limitation).
\begin{equation}
\rho_i\textbf{h}_i^H\textbf{h}_i \ge \rho_j\textbf{h}_j^H\textbf{h}_j \qquad  \forall_{1\le i<j\le N_a}
\label{powerConstraint1}   
\end{equation}

With i.i.d. complex-Gaussian inputs, to find the (instantaneous) mutual information between the input and the output of the point-to-point MIMO channel, the most common approach is expressing the achievable rate in terms of the singular values of the propagation matrix \cite{Rusek2013Scaling}. In (\ref{receivedY}), $\textbf{B}$ is not a i.i.d random matrix due to ordered $\textbf{h}_n$, so we need to normalize $\textbf{B}_n$ by dividing the 2-norm of $\textbf{h}_n$,
\begin{equation}
 \begin{split}
   \textbf{y}&=\textbf{B} \textbf{s}_{\rho} + \textbf{z}\\
    &=\left[ \textbf{P}_1 \otimes \textbf{h}_1, \quad \textbf{P}_2 \otimes \textbf{h}_2, \quad \cdots  ,\quad \right.\\
    &\left. \quad \textbf{P}_N \otimes \textbf{h}_N \right] [\sqrt{\rho_1}\textbf{s}_1^T,  \sqrt{\rho_2}\textbf{s}_2^T,  \cdots,  \sqrt{\rho_N}\textbf{s}_N^T]^T
      + \textbf{z}\\
      &=\left[ \textbf{P}_1 \otimes \frac{\textbf{h}_1}{\|\textbf{h}_1\|_2}, \quad \textbf{P}_2 \otimes \frac{\textbf{h}_2}{\|\textbf{h}_2\|_2},  \quad  \cdots \right.\\
      &\left. \quad \textbf{P}_N \otimes \frac{\textbf{h}_N}{\|\textbf{h}_N\|_2} \right]
       \left[
            \begin{array}{c}
                  \sqrt{\rho}_1 \textbf{s}_1^T \|\textbf{h}_1\|_2 \\
                  \vdots                           \\
                  \sqrt{\rho}_2 \textbf{s}_2^T \|\textbf{h}_2\|_2  \\
                  \vdots \\
                  \sqrt{\rho}_N \textbf{s}_N^T \|\textbf{h}_N\|_2
            \end{array}
      \right]
      + \textbf{z} \\
      & = \tilde{\textbf{B}} \tilde{\textbf{s}} + \textbf{z}
 \end{split}
\end{equation}
where
$$
\tilde{\textbf{B}}=\left[ \textbf{P}_1 \otimes \frac{\textbf{h}_1}{\|\textbf{h}_1\|_2} \quad \textbf{P}_2 \otimes \frac{\textbf{h}_2}{\|\textbf{h}_2\|_2}  \quad  \cdots \quad \textbf{P}_N \otimes \frac{\textbf{h}_N}{\|\textbf{h}_N\|_2} \right]
$$
and
$$
\tilde{\textbf{s}} =[\sqrt{\rho_1} \|\textbf{h}_1\|_2 \textbf{s}_1^T,  \sqrt{\rho_2} \|\textbf{h}_2\|_2 \textbf{s}_2^T,  \cdots,  \sqrt{\rho_N} \|\textbf{h}_N\|_2 \textbf{s}_N^T]^T
$$

$\tilde{\textbf{B}}$ can be regarded as an $MT \times Nd$ random matrix, whose entries are 0-mean and $\frac{1}{MT}$-variance i.i.d. complex Gaussian variables. When the precoding matrices are randomly generated with zero mean for each entry, such approximation shows good accuracy for large $M$.

In the following, we consider the process of demodulation in the $n$-th iteration. We assume that $N_1$ users {\color{black}have been} detected and  $N_2$ out of $N_1$ ($N_2 \le N_1$) users {\color{black}have been} correctly demodulated, so there are $N' = N_1 - N_2$ users stuck in the detected queue yet. The indices of  $N_1$ selected users, $N_2$ demodulated users and $N'$ stuck users form sets $\wedge_{N_1}$, $\wedge_{N_2}$ and $\wedge_{N'}$, respectively. With perfect cancellation performed, the received signal is updated to
\begin{equation}
\widetilde{\textbf{y} }= \textbf{y} - \tilde{\textbf{B}}_{\wedge_{N_2}} \tilde{\textbf{s}}_{\wedge_{N_2}}
\label{receivedY2}
\end{equation}
for $N'$ users,
\begin{equation}
\widetilde{\textbf{y} }= \tilde{\textbf{B}}_{\wedge_{N'} } \tilde{\textbf{s}}_{\wedge_{N'}} + \overline{\textbf{z}}
\label{residualUser}
\end{equation}
where $\overline{\textbf{z}} = \tilde{\textbf{B}}_{\emph{I}_a \setminus \wedge_{N_1}} \tilde{\textbf{s}}_{\emph{I}_a \setminus \wedge_{N_1} } + \textbf{z}$. Then, the the signal to interference plus noise ratio (SINR) of $n$-th  user ($\text{SINR}_n$) can be driven as
\begin{equation}
    \text{SINR}_n = \frac { \frac{1}{ \mathbb{E} \left \{ \frac{1}{\lambda} \right \} }  {\rho}_n \textbf{h}_n^H\textbf{h}_n  }
    {  {\frac{d}{MT}\sum\nolimits_{j=n+1}^{N_a}{ {\rho}_j \textbf{h}_j^H\textbf{h}_j } } + {1} } 
\label{SINR0}
\end{equation}
where $\mathbb{E} \left \{ \frac{1}{\lambda} \right \}$ in the numerator is the mean value of the inverse of $N'd$ eigenvalues of $(\tilde{\textbf{B}}_{\wedge_{N'} }^H \tilde{\textbf{B}}_{\wedge_{N'} })$. The denominator consists of two parts, in which the former is the interference from residual active users, and the latter is the noise power.

Define $\eta:=\frac{N'd}{MT}\in (0,1)$, by asymptotic result in Theorem 2.35 of RMT in \cite{Tulino2004Random} which is also a good approximation for reasonably small-scale matrix dimensions, the eigenvalues of $(\tilde{\textbf{B}}_{\wedge_{N'} }^H \tilde{\textbf{B}}_{\wedge_{N'} })$ have empirical distribution

\begin{equation}
f_{\eta}(x)=\frac{1}{2 \pi \eta x} \sqrt{(x-a)(b-x)}
\end{equation}
for $x\in [a, b]$, and $a=(1-\sqrt{\eta})^2$, $b=(1+\sqrt{\eta})^2$.

With the help of \cite{Gradshtein1980Table}, $\mathbb{E} \left \{ \frac{1}{\lambda} \right \}$ is evaluated as follows
\begin{equation}
\mathbb{E} \left \{ \frac{1}{\lambda} \right \}= \int_{a}^{b}{ \frac{1}{x} f_{\eta}(x) }dx = \frac{MT}{MT-N'd} 
\label{lambda0}
\end{equation}

Inserting (\ref{lambda0}) into (\ref{SINR0}), then
\begin{equation}
 \begin{aligned}
    \text{SINR}_n &= \frac { \frac{(MT-N'd)} {MT} {\rho}_n \textbf{h}_n^H\textbf{h}_n  } {\frac{d}{MT}\sum\limits_{j=n+1}^{N_a}{ {\rho}_j \textbf{h}_j^H\textbf{h}_j  } + 1} \\
 \end{aligned}
\end{equation}

\subsection{users gather into groups}
Owning to massive active users, it becomes too difficult to assign different power to different users. To simplify the calculation, we divide every $K$ active users into a group with the \emph{descending} order of $\textbf{h}^H\textbf{h}$, then the amount of groups $L$ can be easily computed out,
\begin{equation}
    L = \frac{N_a}{K}
\end{equation}

$\textbf{Remark 2}$: Usually an accurate estimate of $N_a$ is not available, but taking into account that it changes slowly, we can regard the estimated results of previous superframe as current $N_a$.

\begin{figure}[!hbt]
    \centering
    \includegraphics[width = 9cm]{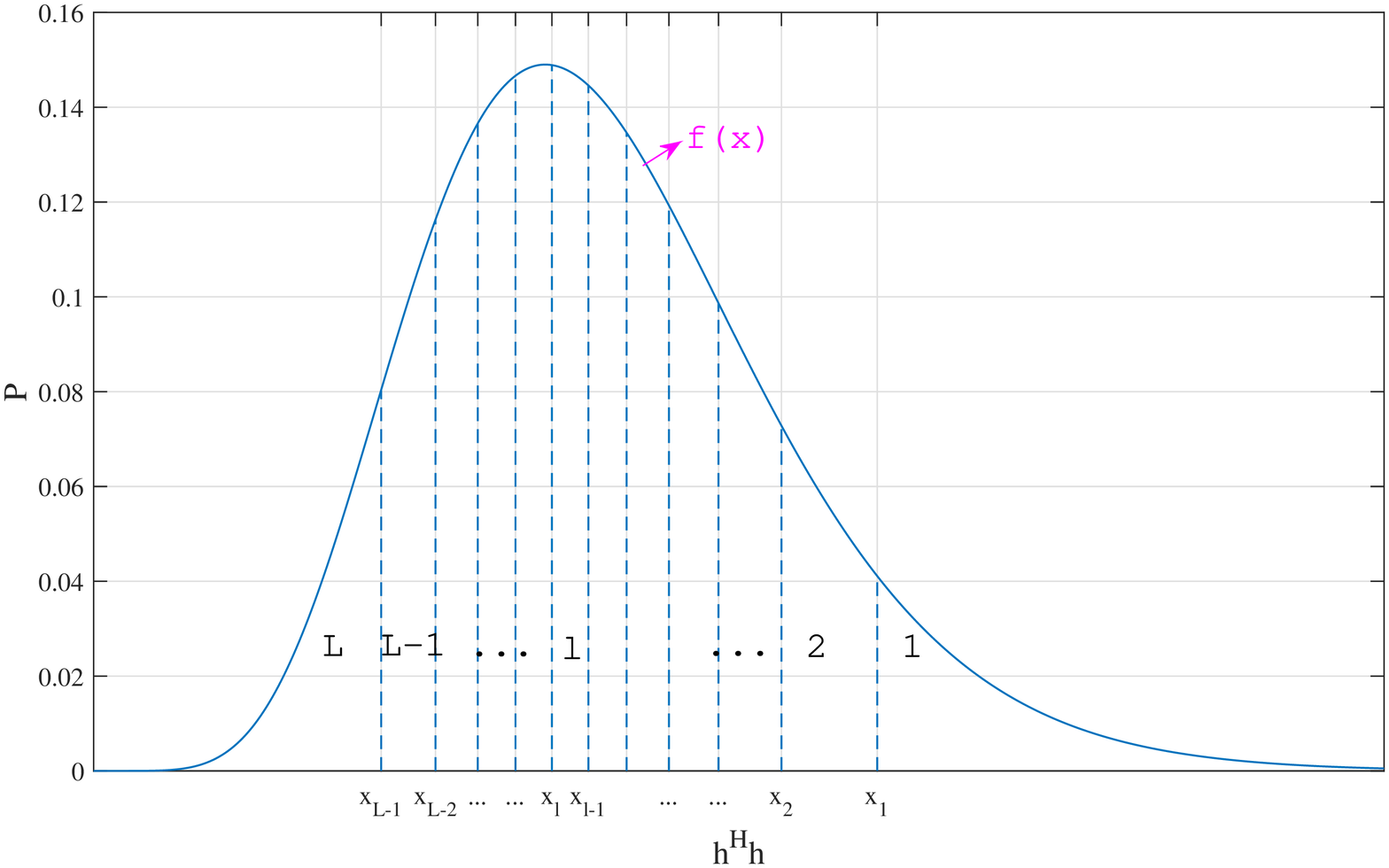}
    \caption{Split points of user grouping on $\textbf{h}^H\textbf{h}$}
    \label{hh_divide}
\end{figure}

Fig.\ref{hh_divide} demonstrates that how a UE confirms its group according to its own  channel amplitude.
Where the split points $x_{l-1}$ and $x_{l}$ for $l$-th group can be calculated by formula (\ref{chiPDF}) and following equation.
\begin{equation}
\int_{x_{l}}^{x_{l-1}} f(x)dx = \frac{1}{L}
\end{equation}
where $l \in [1,L]$, and $x_0 = \infty$, $x_{L} = 0$ for first group and last group, respectively.

Without losing generality, we assume user indices of $l$-th group are $(l-1)K+1$, $(l-1)K+2, \cdots , lK$. Meanwhile, we can use group number $l$ and intra-group number $k$ to represent a certain user as follows.
\begin{equation}
    \begin{aligned}
        U_{(l-1)*K+k} &= U_{l,k}\\
          l\in[1,L] \quad & k \in[1,K]
    \end{aligned}
\label{userindex}   
\end{equation}

Further, we assign users in the same group with the same power $\beta_l$. that is, $\rho_{l,1} = \rho_{l,2} = \cdots = \rho_{l,K}=\beta_l$. Take $l=1$ for an example, the user indices of first group are $1,2,\cdots,K$ and $\rho_{1,1} = \rho_{1,2} = \cdots = \rho_{1,K}=\beta_1$, obviously they are top $K$ strongest users.

Now we consider that the total power of the uplink is limited, i.e
\begin{equation}
    \begin{aligned}
        \sum\limits_{l=1}^{L}{K\beta_l} &= N_a \rho_0\\    
    \end{aligned}
\label{powerConstraint2}   
\end{equation}
where $\rho_0$ is the mean of transmission power.

To keep order statistics, (\ref{powerConstraint1}) becomes
\begin{equation}
    \begin{split}
        \beta_i\textbf{h}_{i,k}^H\textbf{h}_{i,k} &\ge \beta_j\textbf{h}_{j,k'}^H\textbf{h}_{j,k'} \\
        \forall 1\le i<j\le L & \quad k,k' \in [1,K]
    \end{split}
\label{powerConstraint3}
\end{equation}

Further, we assume that $\beta_l=\beta_1 q^{(l-1)}$, ($0<q \le 1$ means stronger users adopt more transmission power). Especially when $q = 10^{-\delta}$, $(\delta \ge 0)$. i.e.,
\begin{equation}
\beta_l=\beta_1 10^{-\delta(l-1)}
\label{powerConstraint4}
\end{equation}
which means there exist gradually degraded with a step of $10\delta$ dB. Based on (\ref{powerConstraint2}) and (\ref{powerConstraint4}), once $\delta$ was known, $\beta_1$ can be calculated out as
\begin{equation}
\beta_1 =
    \begin{cases}
     \frac{L \rho_0 (1-10^{-\delta}) }{ 1-10^{-\delta L} } & {\delta \not= 0 }\\
     1                               & {\delta = 0 }
     \end{cases}
     \label{beta1}
\end{equation}
then we can get $\beta_l$ with (\ref{powerConstraint4}) and (\ref{beta1}).

Assuming perfect SIC is carried out at BS, the SINR of $k$-th uesr in $l$-th group is then
\begin{equation}\label{equsinr}
\begin{split}
&\text{SINR}_{l,k}\\
&= \frac { {\frac{(MT-N'd)}{MT} {\beta}_l \textbf{h}_{l,k}^H\textbf{h}_{l,k} } }
{ \frac{d}{MT}\sum\limits_{l_0=l+1}^{L} \left\{ \beta_{l_0} { \sum_{k_0=k+1}^{K} \textbf{h}_{l_0,k_0}^H\textbf{h}_{l_0,k_0}  } \right\} + 1 }\\
&= \frac { {\frac{(MT-N'd)}{MT} \beta_1 10^{-\delta(l-1)} \textbf{h}_{l,k}^H\textbf{h}_{l,k} } }
{ \frac{d}{MT}\sum\limits_{l_0=l+1}^{L} \left\{ \beta_1 10^{-\delta(l_0-1)} { \sum_{k_0=k+1}^{K} \textbf{h}_{l_0,k_0}^H\textbf{h}_{l_0,k_0} } \right\} + 1 }\\
&\quad  \quad  k\in[1,K] \quad  l\in[1,L]
\end{split}
\end{equation}

Based on the law of large numbers, we can use order statistics $\mathbb{E} \left \{ \textbf{h}_{l_0,k_0}^H\textbf{h}_{l_0,k_0} \right \}$ to replace $\textbf{h}_{l_0,k_0}^H\textbf{h}_{l_0,k_0}$ due to massive connectivity.

The BER of the symbol of interest is given by
\begin{equation}\label{equber}
\text{BER}_{l,k} = Q_1(\text{SINR}_{l,k})
\end{equation}
\begin{equation}
\text{BER} = \frac{\sum_{l=1}^{L} \sum_{k=1}^{K} \text{BER}_{l,k} } {Na}
\end{equation}
and FER yields
\begin{equation}\label{equfer}
\text{FER}_{l,k} = Q_2(\text{SINR}_{l,k})
\end{equation}
\begin{equation}
\text{FER} = \frac{\sum_{l=1}^{L} \sum_{k=1}^{K} \text{FER}_{l,k} } {Na}
\end{equation}
where $Q_1(\cdot)$ or $Q_2(\cdot)$ is a function of $\text{SINR}$ depending on channel coding and modulation mode.

Throughout this paper, analysis of symbol demodulation is based on quadrature phase shift keying (QPSK) modulation scheme. Then
\begin{equation}
\text{BER}_{l,k} = Q(\sqrt{\text{SINR}_{l,k} })
\end{equation}
where Q function is defined as
\begin{equation}
Q(x)=\int_{x}^{\infty}\frac{ e^{-u^2/2} }{ \sqrt{2\pi} }du
\end{equation}

\begin{figure}[!hbt]
    \centering
    \includegraphics[width = 8cm]{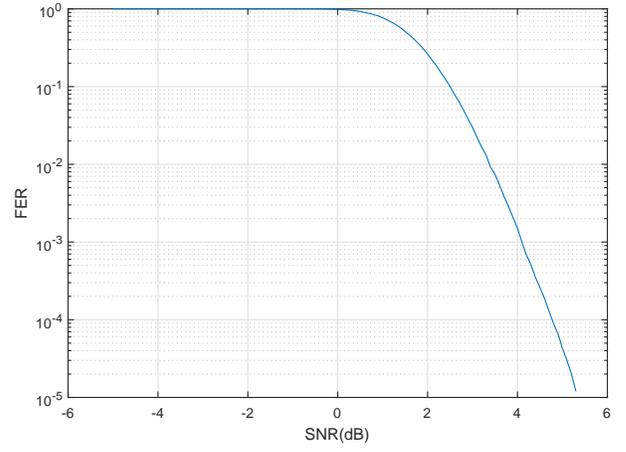}
    \caption{FER-SNR of single user with (2,1,7)Convolutional-QPSK}
    \label{CON_QPSK}
\end{figure}

For massive small packets, FER is an important merit of the quality of communications. We plot a standard FER curve of single user, which can be regarded as $Q_2(\cdot)$ here. See Fig.\ref{CON_QPSK}.

Intuitively, larger $\delta $ will help us to alleviate interference from other active users, at the same time, larger $\delta$ will decrease the power allocated to the users with smaller amplitude of channel gain, thus the performance of such users will be degraded. We should face the question how to select the proper $\delta$ to balance the performance of all active users. Due to the iteration invovled, it is very hard for us to get the closed form of FER and BER  with $\delta$. In this paper, we use Monte Carlo Method to search the optimal $\delta$ with equation \eqref{equsinr}, \eqref{equber} and \eqref{equfer}. In each simulation, we assume users always be detected one by one in {\color{black}descending order of}
$\textbf{h}_n^H\textbf{h}_n$
in the iteration,
see details in the Algorithm \ref{ALGORITHM_FER}, {where $\text{OKrate}[m]$ denotes the probability of successful recovery for the $m$-th user}. Note that $N'$ here could be a decimal, and $\wedge_{N'}$  indicates the set of indices of all stuck users.

\begin{algorithm}[htb]
\caption{\small Theoretical Simulation of FER}
\label{ALGORITHM_FER}
\begin{algorithmic}[1] 
\REQUIRE 
$M$, $N$, $N_a$, $d$, $T$, $K$, $L$, $\delta$, random channel vectors, mean transmission power $\rho_0$, FER-SNR function $Q_2(\cdot)$;
\STATE {Initialization: $\text{FER}[n]=0$, $\text{OKRate}[n]=0$,$1\leq n \leq N_a$};
\STATE {\color{black}calculate} transmission power $\rho_n$ based on $\textbf{h}_n^H\textbf{h}_n$, $1\leq n \leq N_a$; 
\FOR {$n \in \{1,2, \cdots, N_a \} $ }
    \STATE calculate $N_2\Leftarrow \sum_{i=1}^{n}{\text{OKRate}[i]}$
    \STATE calculate $N' = n - N_2 - 1$;
    \FOR {$m \in \wedge_{N'} \cup n $}
        \STATE calculate $\text{SINR}_m$
        $$
        \text{SINR}_m = \frac { {(MT-N'd) \rho_m (\textbf{h}_m^H\textbf{h}_m) } } {{d}\sum_{j=n+1}^{N_a}{ \rho_j (\textbf{h}_j^H\textbf{h}_j ) } + MT }
        $$
        \STATE re-obtain $\text{FER}[m]=Q_2(\text{SINR}_m)$
        \STATE update $\text{OKRate}[m]$
        $$\Leftarrow \text{OKRate}[m] + (1-\text{OKRate}[m]\times (1-\text{FER}[m]) )$$
    \ENDFOR
\ENDFOR
\ENSURE  $\text{FER} = \frac {\sum_{n=1}^{Na}\text{FER}[n]} {N_a}$
\end{algorithmic}
\end{algorithm}

$\textbf{Remark 3}$: In JSPMA, $L$ and $\delta$ are parameters depended on the number of total active users in each transmission. As pointed out before, since the change of traffic load is slowly in  $\text{ms}$ level, which is the duration time of super frame. BS can decide the $L$ and $\delta$ based on last transmission and broadcast these two paramenters  to all online users at the beginning of each super frame. With these two parameter and estimated amplitude of channel gain, active user can decide its transmission power distributely.


\section{NUMERICAL RESULTS}
The simulations study for verifying the proposed scheme are presented in this section. In all simulations, we assume that the BS equipped with $M=8$ antennas and UEs use QPSK for data modulation, and the (2,1,7) convolutional coding is selected as the channel code. Channel vectors are i.i.d complex Gaussian with 0-mean and unit-variance for each element. The $N_a$ active users are chosen uniformly at random among all $N$ online users. We choose information bits of every packet $d=200$ and the frame length $T = 5d$ to be a multiple of the maximum length of short messages. Consequently, there are $8 \times 5 = 40$ shared orthogonal resources.

We make comparison between our proposed JSPMA and simple SPMA in two cases where the amount of active users are 80 and 120 out of 1280 online users, respectively. The number of users per group $K$ is $10$. Accordingly, we get $L = 8$ for 80 active users and $L=12$ for 120 active users.

Note, that different $N_a$ usually corresponds to different best $\delta$. In our simulation, we first we find the best $\delta$ for $N_a=80$ is $0.12$, while $\delta = 0.14$  suits $N_a = 120$ very well, and to ensure the robust stability, we choose $\delta=0.14$ for $N_a = 80$ and $\delta=0.16$ for $N_a = 120$.

To verify our propose algorithm, we choose same $\textbf{P}_n$, and the number of iterations in basic ICBOMP process is set equal to the number of active users $N_a$, i.e., if there are 80 active users out of 1280 online users, the iterations is set to 80. Further, in order to show there is no need to know other users' channel state at UE, the center-controlled scene where the BS sorts all users' channel gain and arranges their power based on their rank is simulated as a reference.

The results show that our proposed algorithm can dramatically enhance our system performance includes BER, FER, UDSR, and overloading factor, especially when $N_a$ get higher.

\begin{figure}[!hbt]
    \centering
    \includegraphics[width = 8cm]{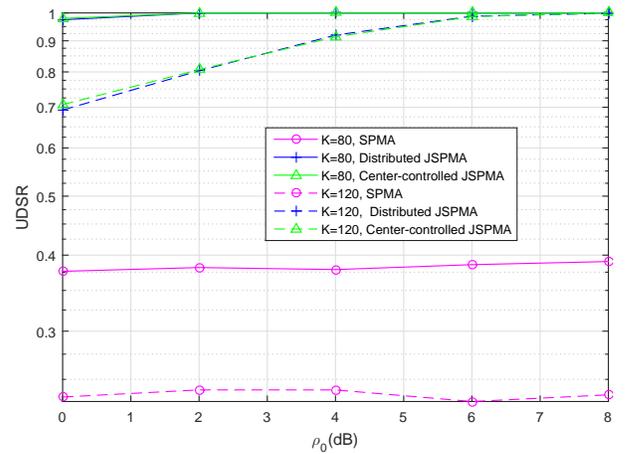}
    \caption{UDSR of SPMA and JSPMA}
    \label{UDSR}
\end{figure}

Test case 1: Fig.\ref{UDSR} shows USDR performances of JSPMA and simple SPMA. The results show that JSPMA improves the performance of UDSR greatly in both $N_a = 80 $ and $N_a = 120$ case. For SPMA, $\text{UDSR}$ is about $40\%$ and $25\%$ with $N_a=80$ and $N_a=120$ respectively in whole region. As a comparison, for JSPMA, when $N_a  = 80$, $\text{UDSR}$ is about $98\%$ even greater than $98\%$ at $\rho_0=0$ dB and then increases to $100\%$ when $\rho_0\geq 2$ dB. When $N_a  = 120$, $\text{UDSR}$ is about $70\%$ at $\rho_0=0$ dB and then increases to $100\%$ when $\rho_0\geq 7$ dB.

\begin{figure}[!hbt]
    \centering
    \includegraphics[width = 8cm]{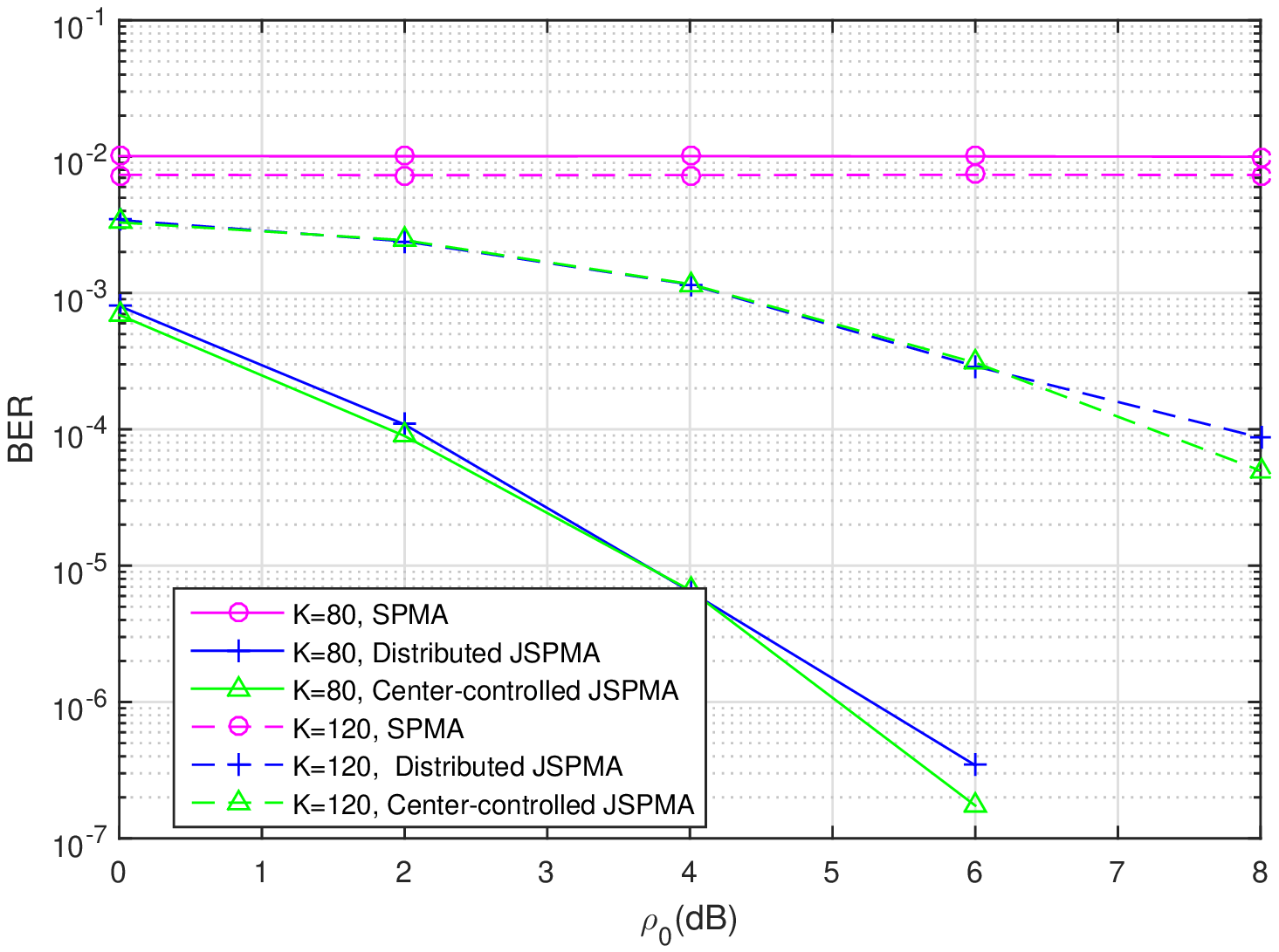}
    \caption{BER of SPMA and JSPMA}
    \label{BER}
\end{figure}

Test Case 2: Fig.\ref{BER} shows that there exist BER error floors about $10^{-2}$ in both cases of $N_a=80$ and $N_a=120$ for SPMA, which indicates that when the amount of active users get higher, the throughput of uplink wireless channel is interference limited, due to the interference among users plays as the key performance-limiting factor. While our proposed algorithm can dramatically improve BER performance with reducing from $10^{-2}$ to $10^{-6}$ at $\rho_0=6$ dB when $N_a=80$, and reaching $10^{-4}$ at $\rho_0=8$ dB when $N_a=120$. The results also demonstrate that JSPMA works very well without knowing others' channel vectors at any UE, the gap between distributed JSPMA and centre controlled JSPMA is as small as 0.5 dB for both cases.

\begin{figure}[!hbt]
    \centering
    \includegraphics[width = 0.51\textwidth]{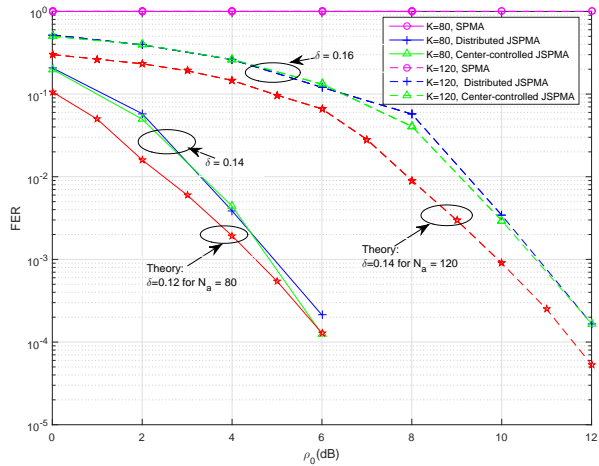}
    \caption{FER of SPMA and JSPMA}
    \label{FER}
\end{figure}

Test Case 3: The FER performances of JSPMA and SPMA are compared in Fig.\ref{FER}. The results show that SPMA can not recover
any transmitted packets. The FER performances are always equal to $1$ in the whole range $[0 ,12]$ {dB} of $\rho_0$ for both cases where $N_a=80$ and $N_a=120$. By contrast, with the help of power assignment, JSPMA shows much more better performance than SPMA. Frame error rate is less than $10^{-3}$ when $N_a=80$ at $\rho_0=5$ dB.
Even for $N_a = 120$, the FER can drop rapidly when $\rho_0$ becomes greater, which proves that our algorithm can get
better effect with the number of active users increasing.



\section{CONCLUSION}
In this paper, we introduce the power domain technique into existing SPMA and propose a new joint domain based massive access for small packets of uplink wireless channel. In the proposed algorithm, users gather into several groups based on their own channel state.
{\color{black}Users in the same group transmit messages with the same power, while there exists a certain degradation between any two adjacent groups.}
BS detects and distinguishes users based on ICBOMP algorithm and then carries out SIC for next detection. The results show that, compared with original SPMA, JSPMA can improve system performance including BER, FER and UDSR greatly. what's more, due to it is a distributed method, the signaling overhead and latency are greatly reduced.
There are $120$ active users simultaneously communicating with BS with $40$ shared orthogonal resources in our system, hence the overloading factor is up to $300\%$. We also compare our distributed method with center-controlled method, the results show that there is only a very narrow gap between two methods.
It is worth noting that our current power assignment is not optimal, we aim to make a balance of power allocation  between strong users and weak users and continue to investigate the  optimal value of $N_a$ and the optimal power $\rho_n$ to guarantee overall system performance.

\end{document}